\documentclass{moriond}

\def\Journal#1#2#3#4{{#1} {\bf #2}, #3 (#4)}

\def\PLB{{\em Phys. Lett.}  B}

\def\PRD{{\em Phys. Rev.} D}

\def\EPJC{{\em Eur. Phys. J.} C}
\def\JHEP{\em JHEP}

\def\ra{\rightarrow}

\def\be{\begin{equation}}
\def\ee{\end{equation}}
\def\bea{\begin{eqnarray}}
\def\eea{\end{eqnarray}}

\def\pt{$p_T$}
\def\dm{$\Delta M$}
\def\ptmiss{$p_T^{miss}$}
\def\el{$\ell$}
\def\tt{$\mathrm{t}\bar{\mathrm{t}}$}

\def\mll{$M(\ell\ell)$}
\def\mtTwo{$m_{T2}(\ell\ell)$}
\newcommand{\Photo}{\includegraphics[height=35mm]{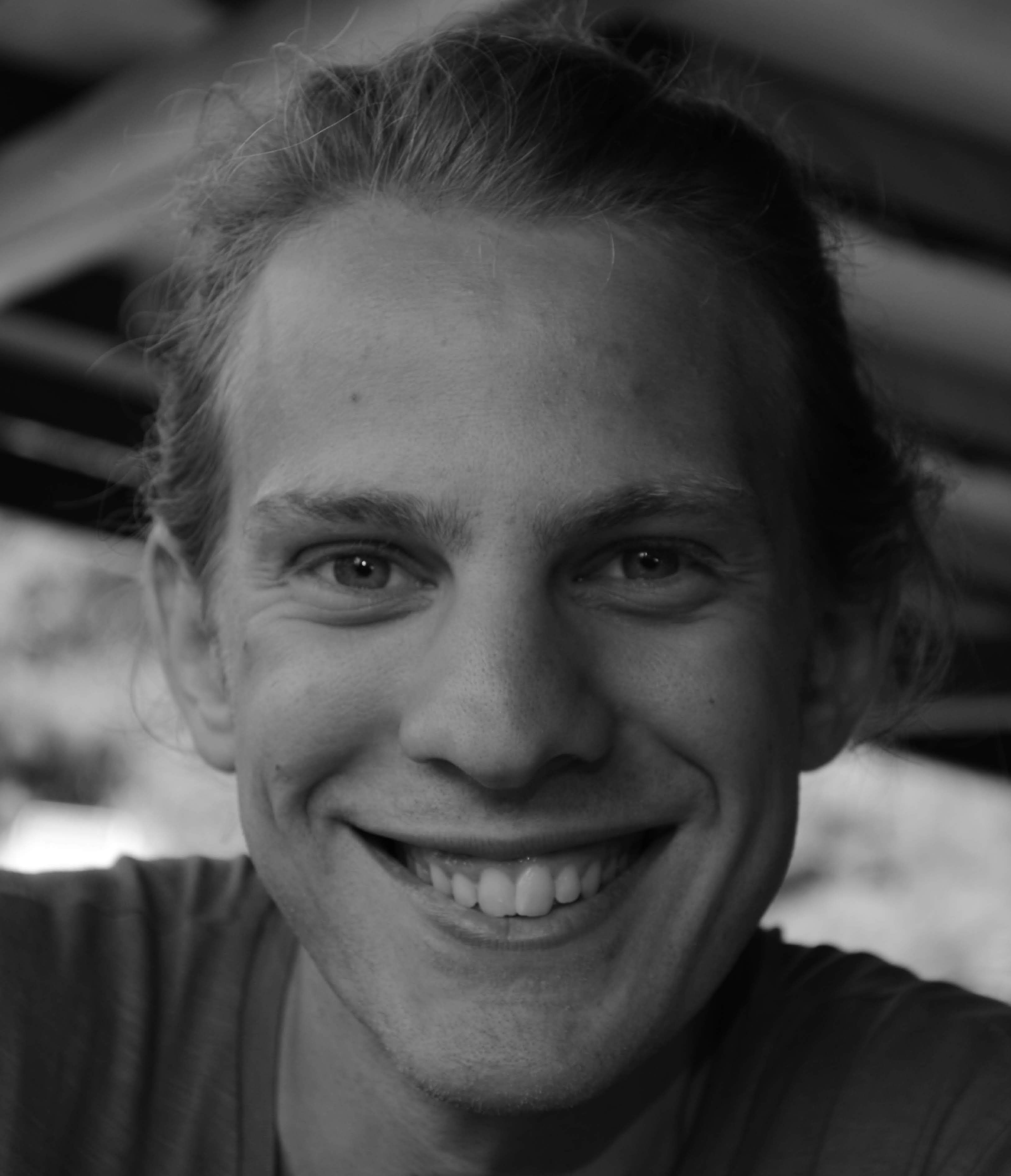}}

\begin{document}
% \linenumbers
\vspace*{4cm}
\title{Searches for Compressed SUSY Models in Leptonic Final States with CMS}

\author{ Emmanouil Vourliotis\\
On behalf of the CMS Collaboration }

\address{National and Kapodistrian University of Athens, Greece}

\maketitle\abstracts{
Searches for supersymmetry (SUSY) models with a compressed mass spectrum are theoretically motivated but also pose experimental challenges. Two recent searches from the CMS Collaboration targeting leptonic final states that can originate from such models are presented. The first search investigates SUSY signatures with two opposite sign or three low momentum leptons, while the second probes the parameter space of top squark models, where the mass difference of the lightest SUSY particles is close to the mass of the top quark. Both searches are based on the full dataset collected by CMS during Run 2 of the Large Hadron Collider, corresponding to 137 fb$^{-1}$.
}

\section{Introduction}

The Standard Model (SM) has been the cornerstone of our understanding of particle physics for the last decades and has proved to be successful in experiments tests consistently. However, there are fundamental questions that it cannot answer, leading to the development of theories beyond it. One of the most promising theories is supersymmetry (SUSY), according to which each SM particle has a SUSY partner with exactly the same properties but different spin. SUSY introduces a new quantum number, $R$-parity, which is different for SM and SUSY particles. If $R$-parity is conserved, as in the models considered below, the lightest SUSY particle (LSP) is stable. The discovery of SUSY particles has been the aim of numerous searches in experiments at the Large Hadron Collider (LHC) throughout its operation.

In recent years, the tide of searches for SUSY has shifted away from signatures with the traditional high transverse momentum (\pt) objects + high missing transverse energy approach. The lack of evidence for SUSY particles in such searches, which has resulted in strong constraints on their masses, has drawn the experimental interest towards SUSY models that can lead to unconventional signatures. An important category of such models are those in which the SUSY particles form a compressed mass spectrum, i.e. the mass difference (\dm) of the LSP and the next-to-LSP (NLSP) is of the order of a few tens of GeV. Such mass differences in the region of the phase space considered lead to final state objects at the edge of detector identification capabilities.

SUSY models with compressed mass spectrum are backed up by strong theoretical motivation. In terms of electroweakino (EWK) production, a scenario where the LSP is a bino and the NLSP is a wino, with small \dm~between them, is motivated by the observed dark matter (DM) relic density~\cite{winobino1,winobino2}, while avoiding constraints by direct DM experiments~\cite{winobino3}. Also in the case of direct higgsino production, naturalness arguments suggest that the lightest higgsinos form a compressed mass spectrum at the electroweak scale~\cite{higgsino}. Finally, in terms of top squark (stop) production, light stops, as the NLSP, forming a compressed mass spectrum with a bino LSP, produce the conditions for their co-annihilation, making the LSP the dominant source of DM~\cite{stop}. From the experimental side, compressed SUSY models pose a number of challenges, such as the identification of low \pt~(soft) objects and the existence of only small amounts of missing transverse energy (\ptmiss), which makes the separation with SM processes difficult. Dedicated analysis methods are usually required to extend the reconstruction acceptance to soft physics objects, as well as to suppress the SM backgrounds.

Recently, the CMS Collaboration has performed two searches targeting compressed SUSY models with leptonic final states. Both of them use the full LHC Run 2 luminosity, 137 fb$^{-1}$, recorded by the CMS experiment at a center of mass energy of $\sqrt{s} = 13$ TeV, and they are presented below.

\section{Search for SUSY in the soft 2\el OS and 3\el~final state}

\begin{figure}[!hbtp]
\centering
\includegraphics[width=0.4\linewidth]{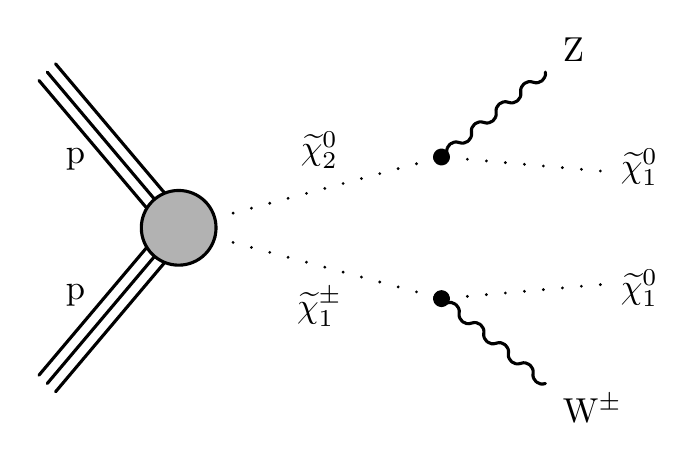}
\includegraphics[width=0.4\linewidth]{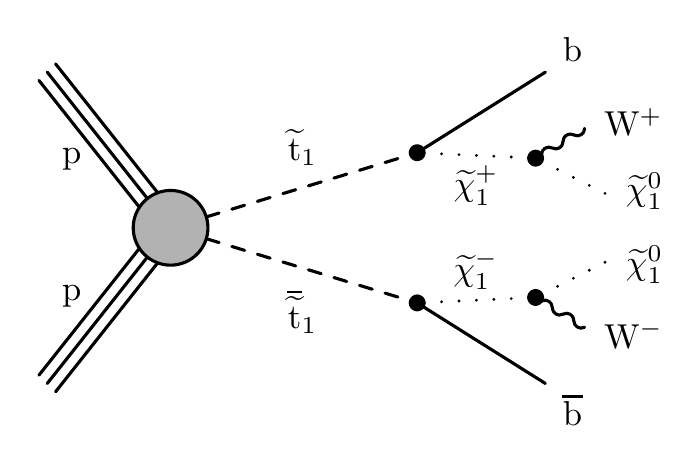}
\caption[]{Example Feynman diagrams of the production and decay of electroweakinos (left) and stops (right).~\cite{sos}}
\label{fig:sos_FD}
\end{figure}

This search~\cite{sos} targets final states containing two or three soft leptons (\el) and moderate to high amount of \ptmiss. Feynman diagrams of typical signal processes for the electroweakino (left) and the stop (right) production are shown in Figure \ref{fig:sos_FD}. In general, a compressed mass spectrum leads to a final state where the objects are of low \pt, due to the small amount of available energy in the decay. In order to boost final state objects and facilitate the separation of the signal from the backgrounds, an initial state radiation (ISR) jet is required. The recoil against the ISR jet induces \ptmiss, which is required to be at least 125 GeV. Events are selected when they contain two opposite sign (OS) leptons, which are soft ($3.5 < p_T < 30$ GeV), isolated and subject to tight impact parameter requirements. For the events of interest, the invariant mass of the lepton pair, \mll, is in the range from 1 to 50 GeV, while low mass resonances are excluded from the selection. The decrease in the lower \mll~ bound is an improvement with respect to the previous result~\cite{oldsos}, resulting in higher acceptance and sensitivity to signal mass hypotheses (masspoints) with even lower \dm.

In the scenario of EWK production, the signal regions (SRs) include two leptons which are required to be of the same flavor (SF), expected to come out of the decay of the off-shell Z boson. Due to this fact, the \mll~has an endpoint at the \dm~between the SUSY particles, making it extremely sensitive for signal and background discrimination. To account for the leptonic decay of the virtual W boson, events with a third soft lepton are allowed in separate SRs. The trilepton final state is a new addition to this analysis and has improved the sensitivity to larger signal \dm, where the acceptance for a soft third lepton increases. In this case, the minimum invariant mass among all the pairs is used as the discriminating variable. In the scenario of stop production, slightly different SRs are designed, where different flavor leptons are allowed as well, since the leptons originate from different decay legs.

The signal modeling is made more accurate in two aspects: Firstly, the computation of the branching fraction of highly off-shell W and Z bosons is corrected as a function of the signal \dm. This is done by taking into account phase space effects at low invariant mass due to the masses of the decay products. Secondly, the hypotheses of different relative sign for mass matrix eigenvalues corresponding to the lightest neutralinos are tested separately, since they affect the \mll~distribution shape.

Owing to the fact that \ptmiss~is a good handle to discriminate the SUSY signal against the SM backgrounds, events are categorized in four \ptmiss~bins. For the trilepton regions, only two \ptmiss~bins are used due to the lower yield of these regions. Events are further split in bins of \mll~in the case of EWK SRs, where the \mll~variable is directly sensitive to the mass splitting of the signal, while the splitting is in bins of the leading lepton \pt~in the case of the stop SRs.

The dominant background in this search is the nonprompt lepton background, i.e. the cases where a jet is misidentified as a lepton or where a lepton from a semileptonic decay of a heavy flavor hadron is selected. This source of background becomes increasingly important and hard to estimate the lower \pt~the selected leptons have. It is estimated using a data-driven fake rate method. A dedicated control region (CR) with the same selection criteria as the signal selection but including same sign leptons and, therefore, dominated with nonprompt leptons, is used to further constrain the nonprompt lepton prediction uncertainty from data. Concerning  prompt background processes, the major ones are the dilepton \tt, which is reduced by a b jet veto, and the DY$\ra\tau\tau$, which is reduced by excluding events with the approximate mass of the Z boson, reconstructed from the final leptons and the \ptmiss, in the range from 0 to 160 GeV. Both process are normalized to data in high purity CRs, designed by inverting the respective requirement used to reduce the contribution of each process. The WZ$\ra 3 \ell$ process is the most important prompt background in the trilepton regions, even though it is reduced by vetoing the Z peak in the \mll~distribution. It is estimated from simulation and then normalized to data in a specially designed CR. All other diboson processes are effectively vetoed by an upper bound to the transverse mass of the leptons and the \ptmiss~($m_T(\ell_i,p_T^{miss}) < 70$ GeV) and are estimated purely from simulation. The latter requirement is not applied to the stop SRs to increase the signal acceptance.

\begin{figure}[!hbtp]
\centering
\includegraphics[width=0.32\linewidth]{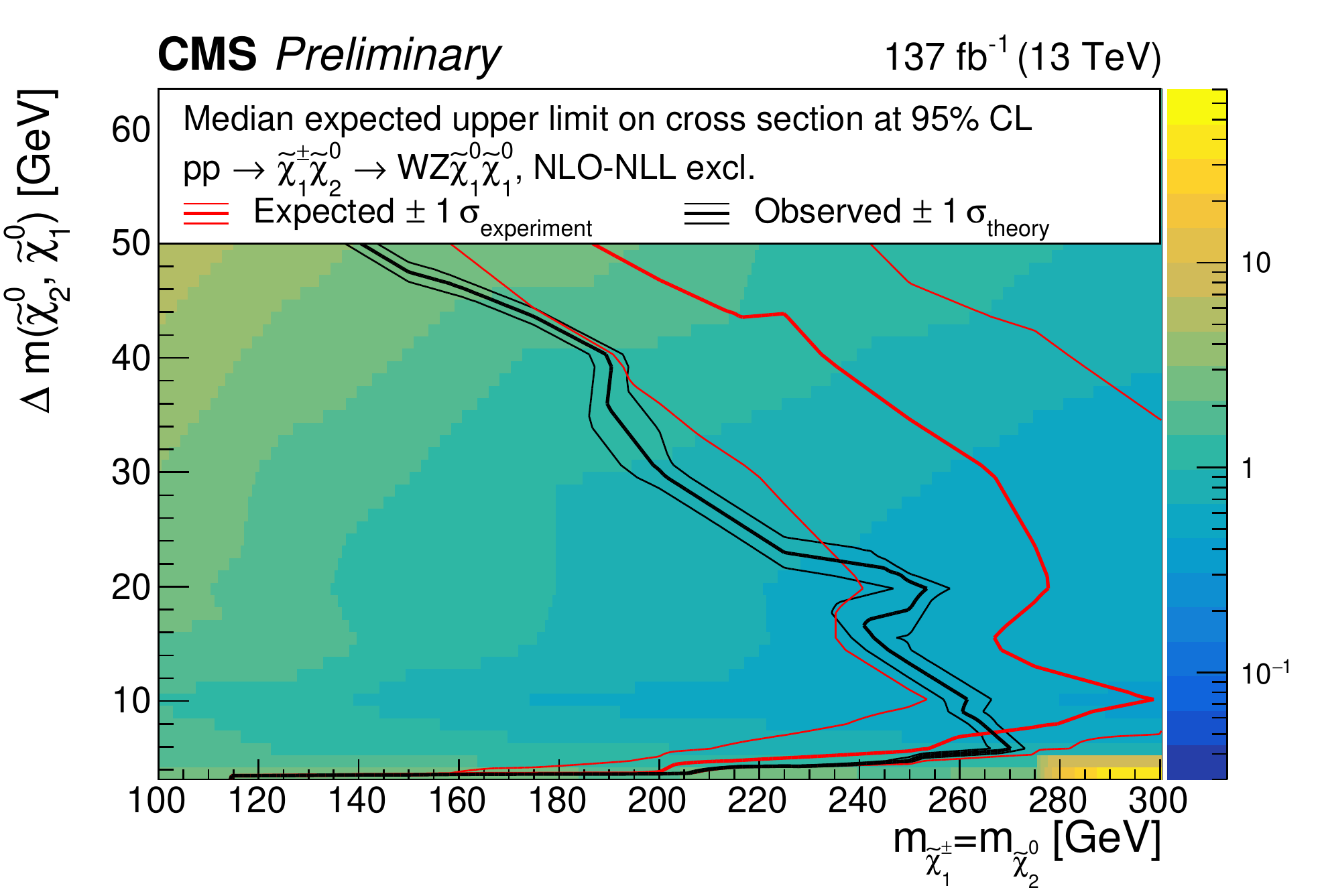}
\includegraphics[width=0.32\linewidth]{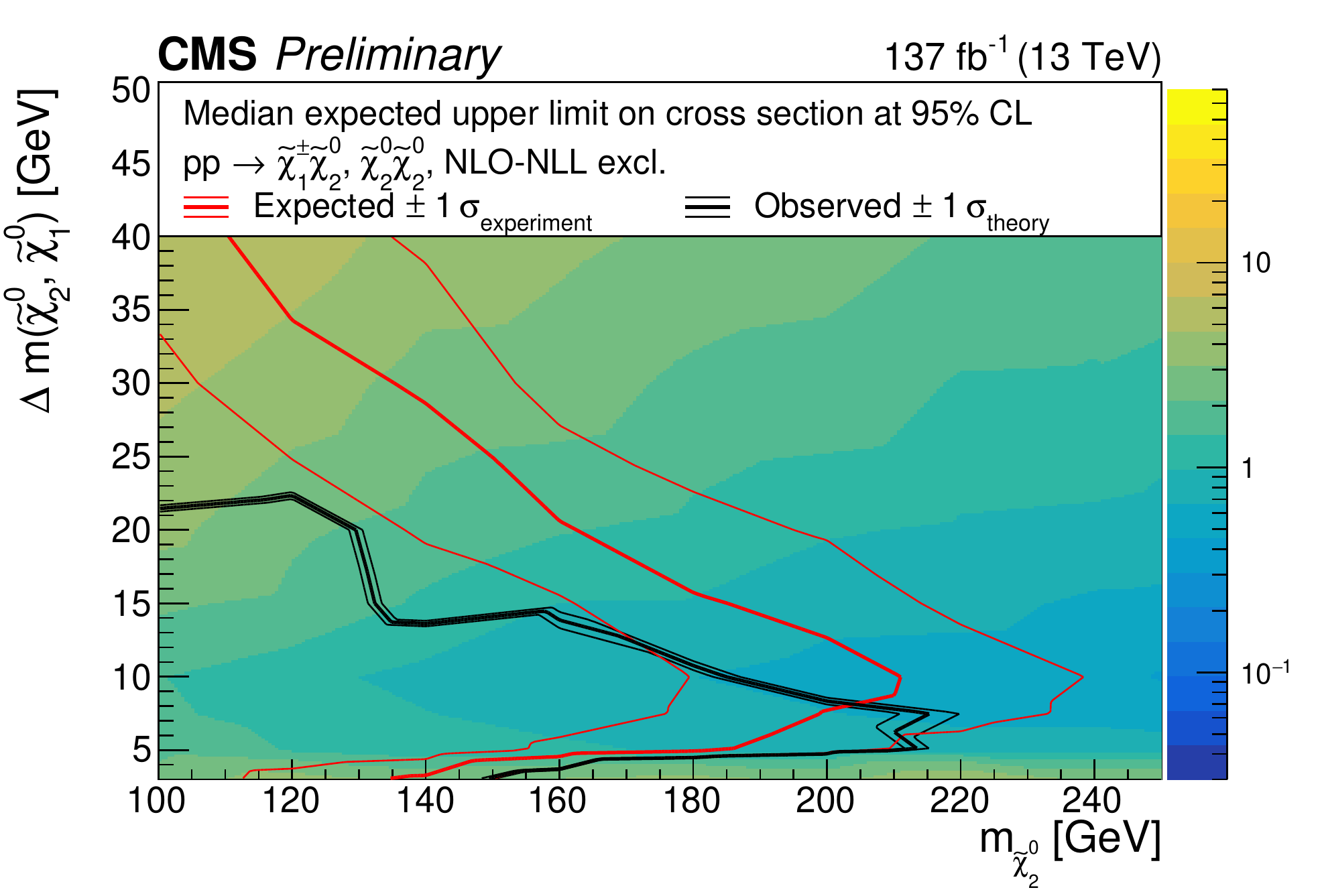}
\includegraphics[width=0.32\linewidth]{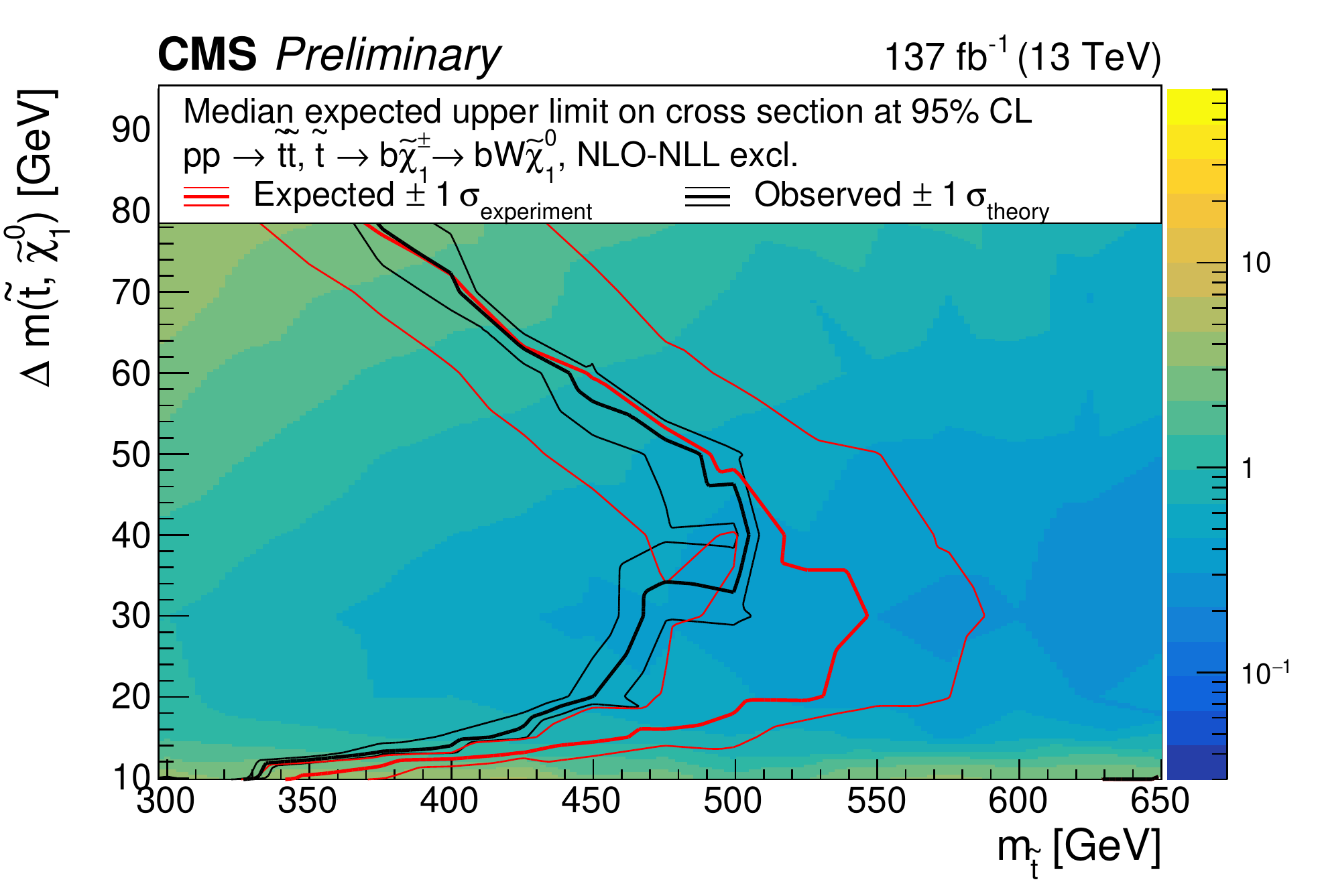}
\caption[]{Exclusion limits at 95\% CL for a wino/bino simplified model (left), a higgsino simplified model (center) and a stop simplified model in the case of chargino mediated decay (right).~\cite{sos}}
\label{fig:sos_limits}
\end{figure}

No significant excesses of the observed data yields are found with respect to the background prediction. The results are interpreted in terms of exclusion limits to a wino/bino simplified model (Figure \ref{fig:sos_limits}, left), a higgsino simplified model (Figure \ref{fig:sos_limits}, center), a pMSSM inspired model with the higgsino as the LSP and stop production simplified models, one including the four body decay of the stop and the other including the chargino mediated decay (Figure \ref{fig:sos_limits}, right). These limits constitute a significant improvement with respect to the previous CMS analysis in this final state~\cite{oldsos} and showcase the increased sensitivity in the very low \dm~region. In the case of the wino/bino simplified model, the sensitivity reaches up to $m_{\tilde{\chi}_2^0} = 300$ GeV for $\Delta M \approx 10$ GeV, while in the case of the higgsino simplified model it reaches up to $m_{\tilde{\chi}_2^0} = 215$ GeV for the same mass splitting. Thanks to the increased acceptance and the strategy optimization, very low \dm, down to 3 GeV, can be probed for both models.

\section{Search for light stop pairs in 2\el~final state (top corridor)}

The region of SUSY parameter space where the mass splitting between the NLSP stop and the LSP neutralino is close to the mass of the SM top quark ($|m_t - ( m_{\tilde{\mathrm{t}}} - m_{\tilde{\chi}_1^0})| < 30$ GeV) and the LSP is light ($m_{\tilde{\chi}_1^0}$ up to $100$ GeV) is called ``top corridor". In the top corridor, the SUSY signal signature mimics the SM top pair production, therefore requiring dedicated techniques to be probed. CMS targets this corner of the parameter space with the specially designed search~\cite{topcorridor} described below.

The event selection follows the standard selection for the \tt~process: two opposite-sign light leptons (e, $\mu$) with $p_T > 35$ GeV are required, along with at least two jets, at least one of which is a b jet. In contrast with the SM \tt~production, the SUSY signal includes also a pair of LSPs in the final state, which induce low to high amounts of \ptmiss. To benefit from this difference, the analysis selects events with $p_T^{miss} > 50$ GeV. The existence of the LSPs in the signal final states also negates the kinematic endpoint at the mass of the W boson of the stransverse mass of the leptons, \mtTwo, defined as

\begin{equation}
    m_{T2}(\ell\ell) = \min_{\vec{p}_{T,1}^{~miss} + \vec{p}_{T,2}^{~miss} = \vec{p}_{\mathrm{T}}^{\mathrm{~miss}}} \left( \max \left[ m_T(\vec{p}_T^{~\ell 1},\vec{p}_{T,1}^{~miss}) , m_T(\vec{p}_T^{~\ell 2},\vec{p}_{T,2}^{~miss}) \right] \right)\mathrm{,}
\end{equation}

\noindent which is present for the \tt~and tW SM processes. This provides another handle to separate the signal from the background by imposing $m_{T2}(\ell\ell) > 80$ GeV. Finally, the contribution from SM processes is further reduced by vetoing the Z boson peak and all the low mass resonances for the OS and SF pairs.

The selected events are split in nine SRs, one for each year (2016, 2017, 2018) and one for each flavor combination (ee, $\mu\mu$, e$\mu$). The dominant background process is the SM \tt~process, contributing more than $90\%$ in the SRs. Due to the requirements on \mtTwo~and \ptmiss, the \tt~contribution mostly comes from events with mismeasured jets or, less frequently, with nonprompt leptons. The estimation of this background is performed by using the simulation with the best CMS measurements of inclusive and differential cross section to date~\cite{top1,top2}. Given its place as the main background for the analysis, multiple modeling uncertainties are taken into account for the \tt~process. These include uncertainties for the choice of PDFs and $\alpha_S$, for the modeling of inital and final state radiation, of the top quark \pt~spectrum and  of the matrix element/parton shower merging, as well as for the variation of the top quark mass. The rest of the background comes from the SM tW process ($\sim 4\%$) and DY, VV or \tt V processes ($\sim 6\%$). These minor background contributions are also estimated from simulation.

\begin{figure}[!hbtp]
\centering
\includegraphics[width=0.32\linewidth]{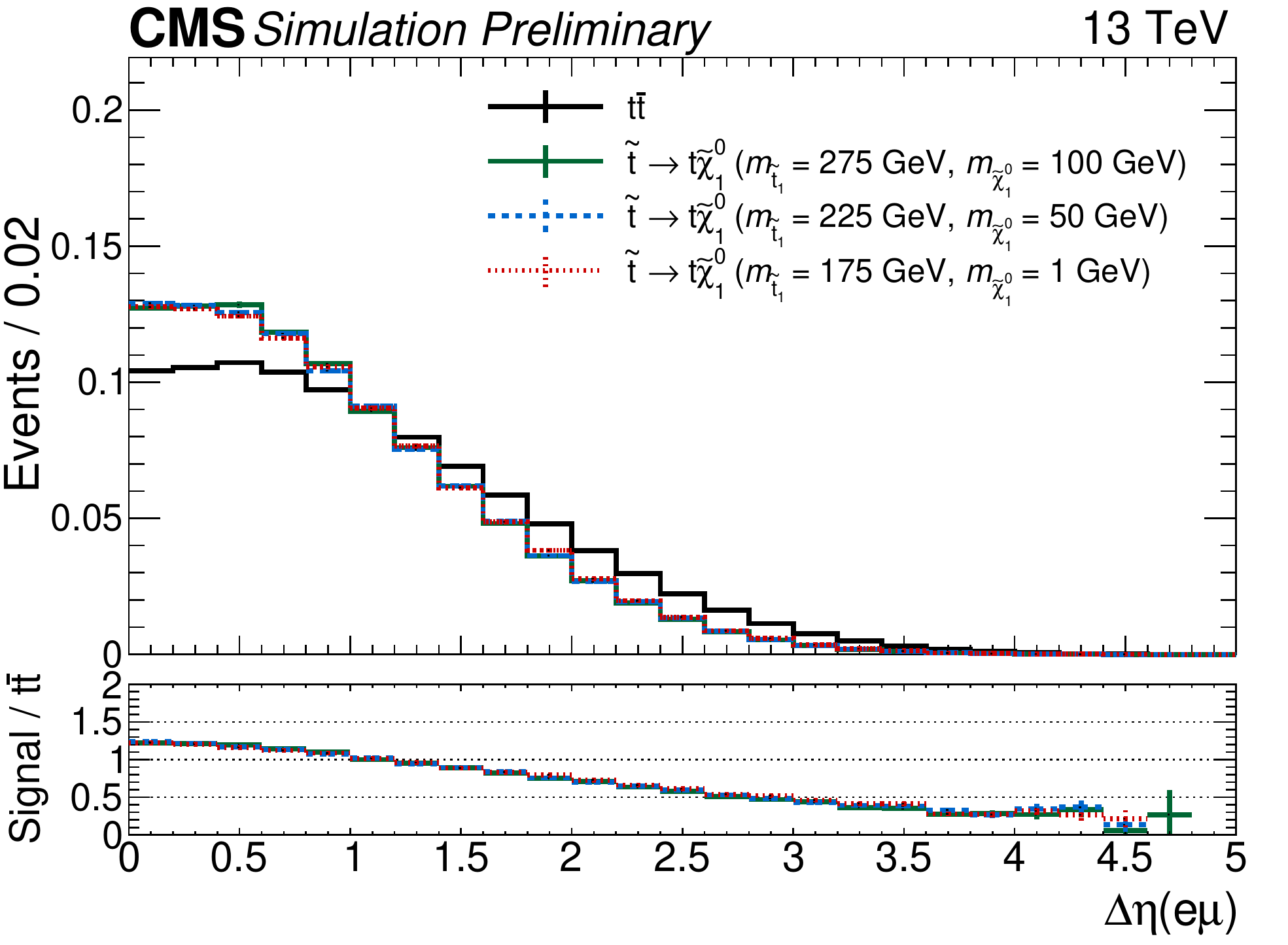}
\includegraphics[width=0.32\linewidth]{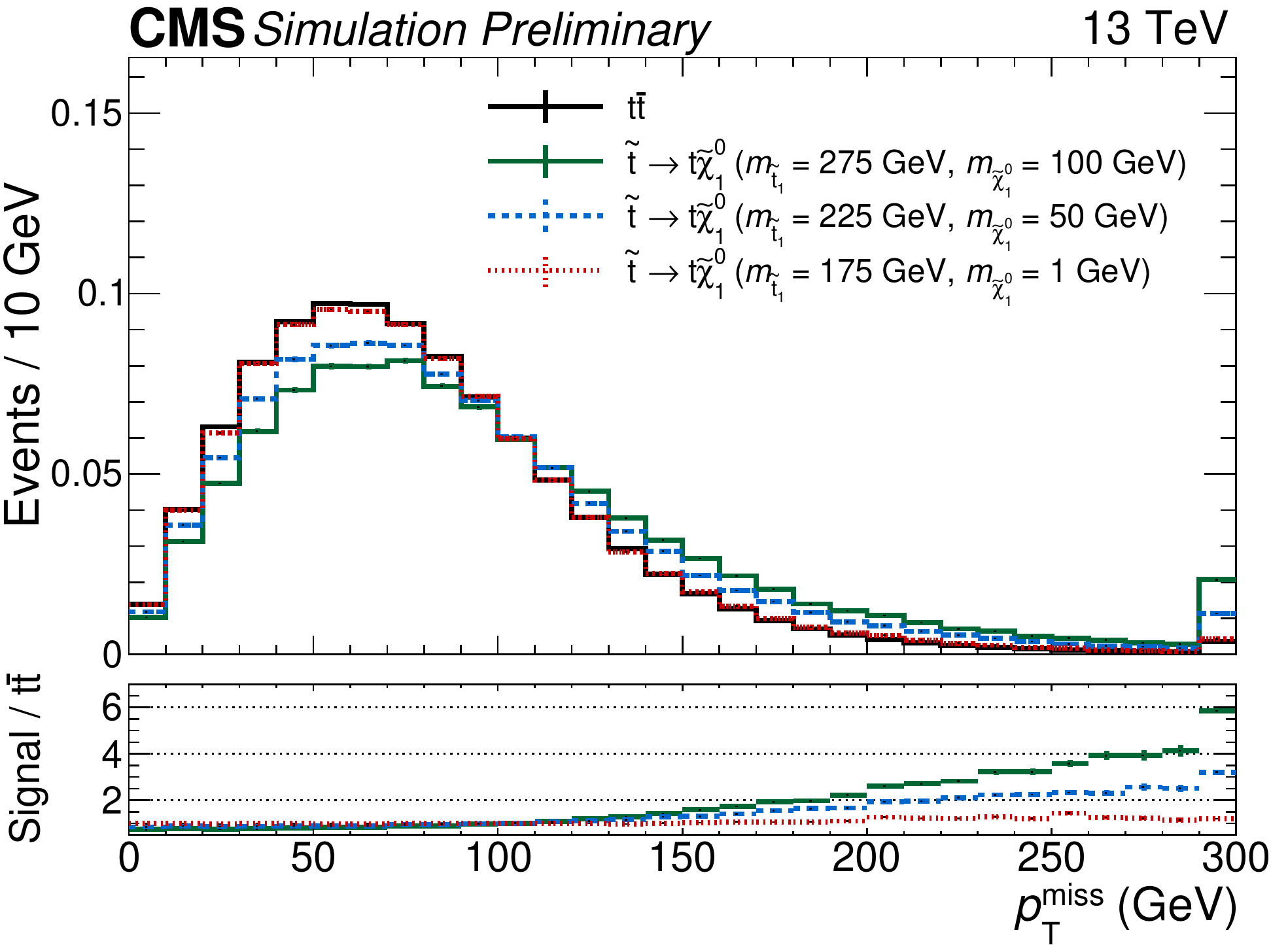}
\includegraphics[width=0.32\linewidth]{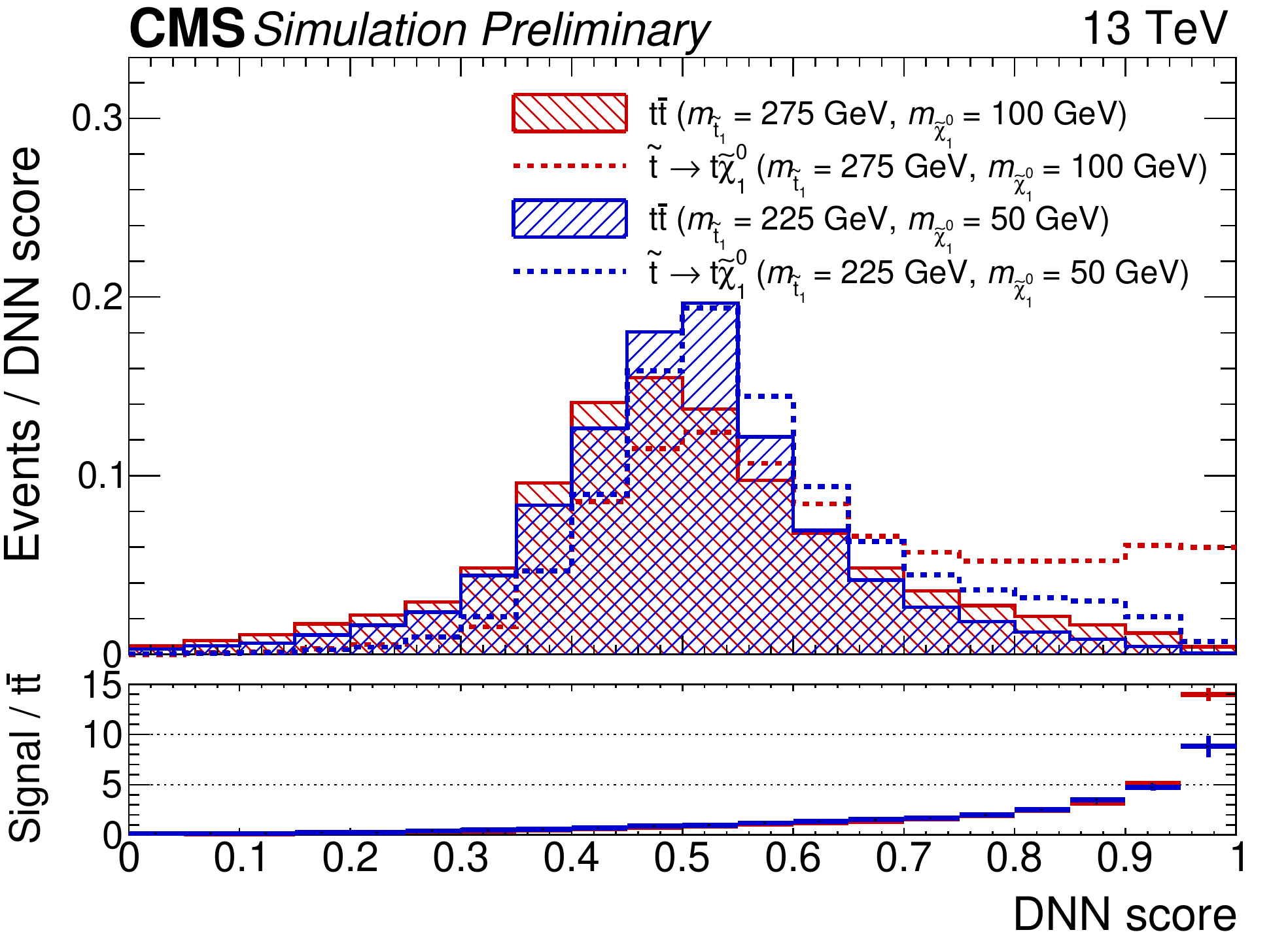}
\caption[]{Examples of input variables for the DNN (left: $\Delta\eta(\ell\ell)$, center: \ptmiss) for the background and different signal masspoints. Example output of the DNN for two different signal masspoints (right).~\cite{topcorridor}}
\label{fig:topcorr_DNN}
\end{figure}

The similarity of the SUSY signal signature to the one of the SM \tt~production makes their separation challenging. In order to tackle this problem and maximize the sensitivity of the search, a multivariate technique is utilized. More specifically, a deep neural network (DNN) is designed for the discrimination of the signal and the background. The DNN is 7 layers deep and takes as input event-level variables (\mtTwo, \mll, \ptmiss, $H_T$), single lepton variables (\pt, $\eta$ for each lepton) and dilepton variables ($p_T(\ell\ell)$, $\Delta\eta(\ell\ell)$, $\Delta\phi(\ell\ell)$). Examples of the input variables for the background and different signal masspoints are shown in Figure \ref{fig:topcorr_DNN} (left, center). From the two left plots of Figure \ref{fig:topcorr_DNN}, it is evident that different signal masspoints have different separation from the background, which would call for different DNN trainings for each masspoint in order to maximize the sensitivity gain.

The novel technique that this analysis uses lies in setting up the DNN training in such a way, so that a single DNN gets all the information from the different signal masspoints and produces optimized weights for all of them. This is called \emph{parametric training}~\cite{paramTrain} and is implemented by including the mass of the stop ($m_{\tilde{\mathrm{t}}}$) and the mass of the LSP ($m_{\tilde{\chi}_{1}^{0}}$) as parameters in the training. These parameters do not contribute in the signal-background separation but rather regulate how all the other input variables are used by the DNN in order to achieve the best result for any given parameter value. The output of the parametric DNN for two example signal masspoints is shown in Figure \ref{fig:topcorr_DNN} (right), which presents the great discrimination power that this single DNN maintains for very different combinations of $m_{\tilde{\mathrm{t}}}$ and $m_{\tilde{\chi}_{1}^{0}}$. While the reinterpretation of the CMS \tt~inclusive cross section measurement~\cite{top1} as a search for extra particles on top of the SM \tt~production and the previous CMS search in this final state~\cite{oldtopcorridor} provided sensitivity to regions with very low $m_{\tilde{\chi}_{1}^{0}}$ or regions with $m_{\tilde{\mathrm{t}}} - m_{\tilde{\chi}_{1}^{0}} \approx m_\mathrm{t}$, this search extends the sensitivity to regions with higher values of $m_{\tilde{\chi}_{1}^{0}}$ and $m_{\tilde{\mathrm{t}}} - m_{\tilde{\chi}_{1}^{0}}$.

% The improvements in sensitivity with respect to the previous \tt~inclusive cross section measurement~\cite{top1} and the previous iteration of this analysis~\cite{oldtopcorridor} are larger in the regions of high $m_{\tilde{\chi}_{1}^{0}}$ and/or regions with ($m_{\tilde{\mathrm{t}}} - m_{\tilde{\chi}_{1}^{0}}$) away from the $m_\mathrm{t}$.

\begin{figure}[!hbtp]
\centering
\includegraphics[width=0.4\linewidth]{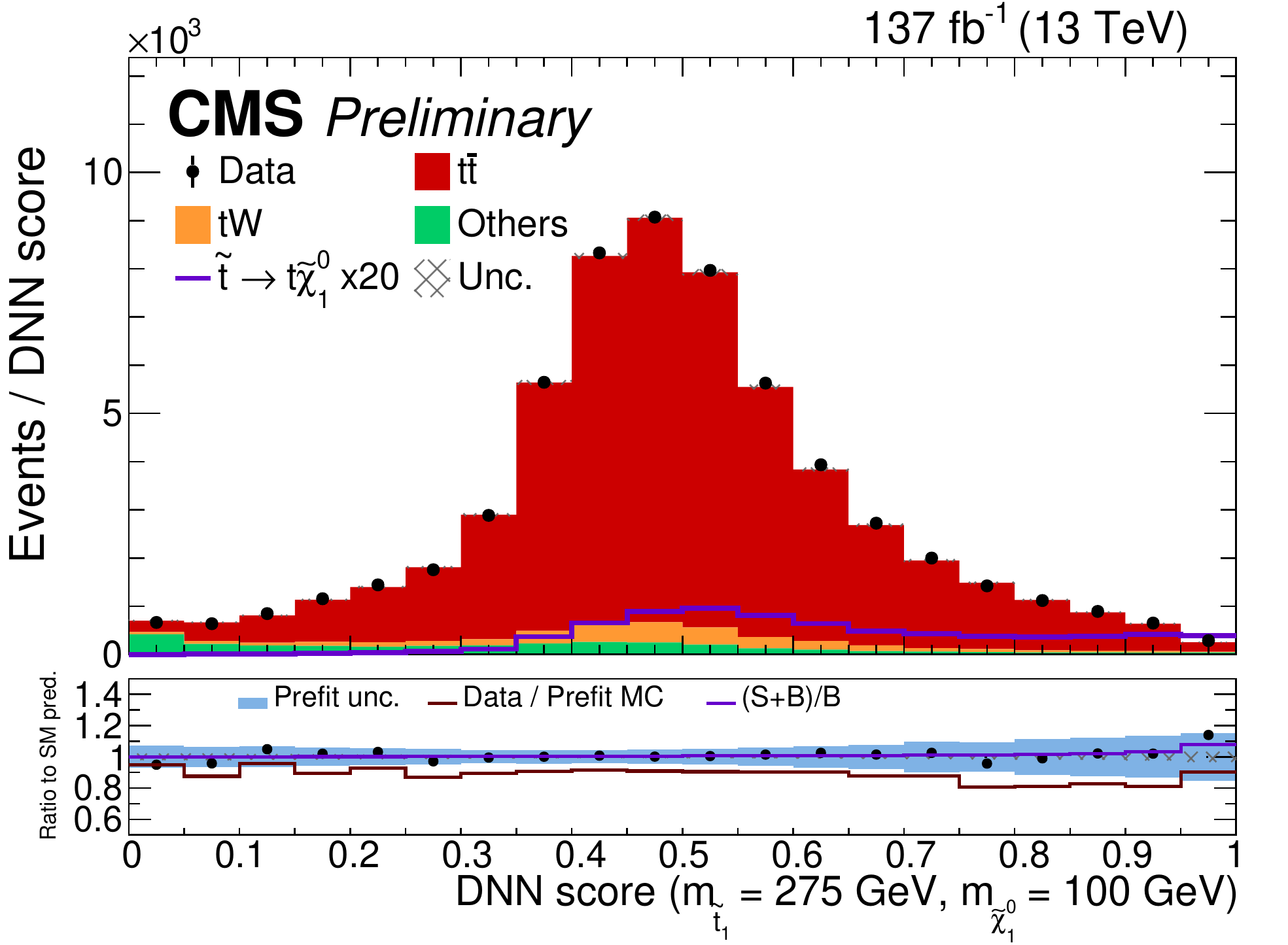}
\includegraphics[width=0.42\linewidth]{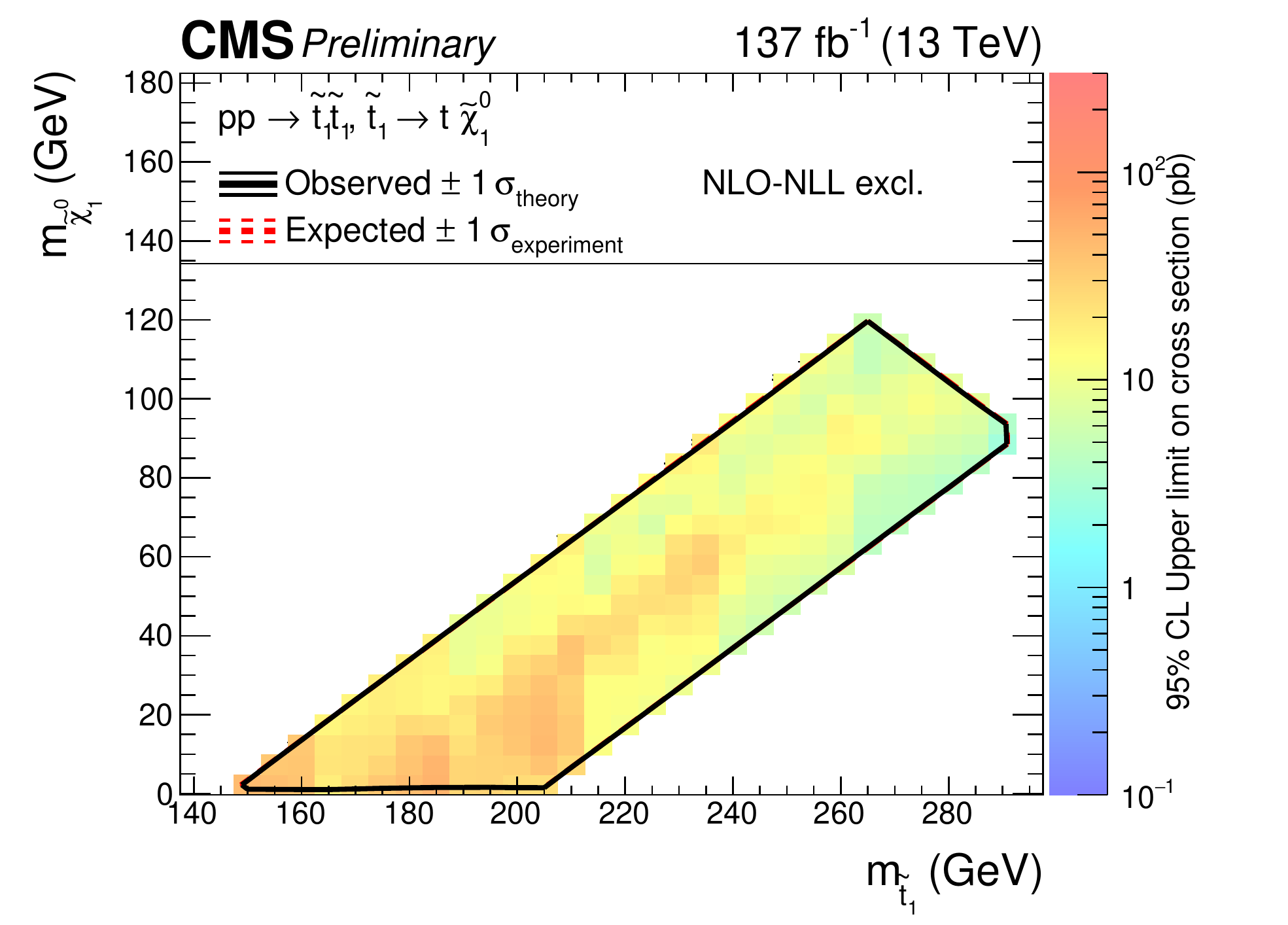}
\caption[]{Post-fit DNN score distribution in the signal region for the mass hypothesis $m_{\tilde{\mathrm{t}}}=275$ GeV and $m_{\tilde{\chi}_{1}^{0}}=100$ GeV (left) and exclusion limits at 95\% CL for stop production in the top corridor (right).~\cite{topcorridor}}
\label{fig:topcorr_results}
\end{figure}

Good agreement between data and simulated background is observed in all the SRs and for all signal masspoints, as it can be seen in Figure \ref{fig:topcorr_results} (left) for the signal mass hypothesis $m_{\tilde{\mathrm{t}}}=275$ GeV and $m_{\tilde{\chi}_{1}^{0}}=100$ GeV. The analysis sets exclusion limits at $95\%$ confidence level (CL) on the cross section as a function of the stop mass and the LSP mass, shown in Figure \ref{fig:topcorr_results} (right). For the first time in CMS, the whole region of the top corridor is excluded.

\section{Summary and Future Prospects}

The CMS Collaboration is actively investigating compressed SUSY models, successfully covering the less explored corners of the SUSY parameter space. These models have very strong theoretical motivations but are difficult to probe experimentally. To overcome the experimental challenges, new ideas and techniques are designed and implemented, expanding the sensitivity limits. Looking to the future, these limits will benefit to a great extent from experiment upgrades~\cite{BSMinHLLHC} and improvements to analysis methods.

\end{document}